\documentclass[twocolumn,aps,superscriptaddress,showpacs]{revtex4}
\usepackage{amssymb}
\usepackage{amsmath}
\usepackage{graphicx}
\usepackage[normalem]{ulem}
\usepackage[dvips]{color}

\begin{document}

\title{Higher-order anisotropic flows and dihadron correlations in
Pb-Pb collisions at $\sqrt{s_{NN}}=2.76$ TeV in a multiphase transport model}
\author{Jun Xu}\email{xujun@comp.tamu.edu}
\affiliation{Cyclotron Institute, Texas A\&M University, College
Station, Texas 77843-3366, USA}
\author{Che Ming Ko}\email{ko@comp.tamu.edu}
\affiliation{Cyclotron Institute and Department of Physics and
Astronomy, Texas A\&M University, College Station, Texas 77843-3366, USA}

\date{\today}

\begin{abstract}
Using a multiphase transport model that includes both initial
partonic and final hadronic scatterings, we have studied
higher-order anisotropic flows as well as dihadron correlations as
functions of pseudorapidity and azimuthal angular differences
between trigger and associated particles in Pb-Pb collisions at
$\sqrt{s_{NN}}=2.76$ TeV. With parameters in the model determined
previously from fitting the measured multiplicity density of
mid-pseudorapidity charged particles in central collisions and their
elliptic flow in mid-central collisions, the calculated higher-order
anisotropic flows from the two-particle cumulant method reproduce
approximately those measured by the ALICE Collaboration, except at
small centralities where they are slightly overestimated. Similar to
experimental results, the two-dimensional dihadron correlations at
most central collisions show a ridge structure at the near side and
a broad structure at the away side. The short- and long-range
dihadron azimuthal correlations, corresponding to small and large
pseudorapidity differences, respectively, are studied for triggering
particles with different transverse momenta and are found to be
qualitatively consistent with experimental results from the CMS
Collaboration. The relation between the short-range and long-range
dihadron correlations with that induced by back-to-back jet pairs
produced from initial hard collisions is also discussed.
\end{abstract}

\pacs{25.75.-q, %Relativistic heavy-ion collisions
      12.38.Mh, %Quark-gluon plasma
      24.10.Lx  %Monte Carlo simulations (including hadron and parton
                %cascades and string breaking models)
      }

\maketitle

\section{Introduction}
\label{introduction}

Heavy ion collisions at the Relativistic Heavy-Ion Collider (RHIC)
have made it possible to obtain important information on the
properties of the quark-gluon plasma (QGP) that is produced in these
collisions~\cite{Ars05,Adc05,Bac05,Ada05}. Among the many
observables that have been used to infer the properties of the QGP
are the anisotropic flows of produced particles, particularly the
elliptic flow ($v_2$)~\cite{Son11} that corresponds to the
second-order harmonic coefficient of the anisotropic flow and is
appreciable in noncentral collisions of identical
nuclei. More recently, it was found that the
triangular flow $v_3$ that corresponds to the third-order harmonic
coefficient of the anisotropic flow could also be significant in
collisions of all centralities as a result of initial density
fluctuations in the collision geometry~\cite{Alv10,Sor10}.
Furthermore, it was shown that studying $v_3$ together with $v_2$
can help better constrain the properties of the
QGP~\cite{Sch11,Ada11}. Also, it was suggested that useful
information about the QGP could be obtained from the dihadron
correlation~\cite{Ada05b} between lower-$p_T$ associated particles
from the medium response to the away-side jet and higher-$p_T$
trigger particles produced from the near-side jet. Studies based on
the transport model showed that higher-order anisotropic flows,
especially the triangular flow, had a large effect on dihadron
correlations induced by initial energetic jets~\cite{Ma11,Xu11a}. To
extract from the dihadron correlation information on the
interactions of energetic jets produced from initial hard collisions
with medium partons in the quark-gluon plasma requires, however, the
subtraction of the background contribution from anisotropic flows
and their fluctuations, and this is a topic of great current
interest~\cite{Alv10,Ma11,Xu11a,Aga10,Luz11}.

Recently, results from the first Pb-Pb collisions at
$\sqrt{s_{NN}^{}}=2.76$ TeV at the Large Hadron Collider (LHC) have
attracted a lot of attentions~\cite{Aam10a,Aam10b,Aam10c,Aam10d}. In
the most central collisions ($0-5\%$), the multiplicity density of
produced charged particles at mid-pseudorapidity is $2.2$ times of
that in Au+Au collisions at $\sqrt{s_{NN}^{}}=200$ GeV at RHIC. The
measured elliptic flow, which can provide more reliable information
about the produced QGP~\cite{Nie11}, is of a similar magnitude as
that measured at RHIC. Besides, experimental data on higher-order
flows~\cite{Aam11} and dihadron correlations~\cite{Cha11} have also
become available, providing the opportunity to study the properties
of the hotter and denser QGP formed at LHC. In particular, it was
found that the long-range dihadron azimuthal correlations could be
entirely accounted for by the anisotropic flows~\cite{Aam11,Cha11}.

In a recent work~\cite{Xu11b}, we studied Pb-Pb collisions at
$\sqrt{s_{NN}^{}}=2.76$ TeV in a multiphase transport (AMPT) model.
The AMPT model is a hybrid model with the initial particle
distribution generated by the heavy ion jet interaction generator
(HIJING) model~\cite{Xnw91}. In the version of string melting, which
was used in the previous work~\cite{Xu11b} and will be used in the
present study, all hadrons produced in the HIJING model through Lund
string fragmentation are converted to their valence quarks and
antiquarks, whose evolution in time and space is modeled by Zhang's
parton cascade (ZPC) model~\cite{Zha98}. After their scatterings,
quarks and aniquarks are converted via a spatial coalescence model
to hadrons, and the scatterings among them are described by a
relativistic transport (ART) model~\cite{Li95}. For a recent review
of the AMPT model, we refer the readers to Ref.~\cite{Lin05}. By
adjusting the values of parameters in the Lund string fragmentation
function and the parton scattering cross section, we fitted the
differential elliptic flow at $40-50\%$ centrality window and
reproduced reasonably well the centrality dependence of both the
multiplicity density and the elliptic flow of mid-pseudorapidity
charged particles~\cite{Xu11b}, although the transverse momentum
spectrum at most central collisions is softer than the data from the
ALICE Collaboration as a result of the use of massless partons in
the model that are less affected by the radial flow~\cite{Lin05}. In
the present paper, we study higher-order anisotropic flows and
dihadron correlations in Pb-Pb collisions at LHC by using the AMPT
model with the newly fitted parameters. Our results show that in
heavy ion collisions at LHC higher-order flows are appreciable and a
ridge structure exists in the dihadron correlation as seen in the
experiments~\cite{Cha11}. The ridge structure is a result of the
longitudinal expansion of the anisotropic flows as pointed out
previously in Ref.~\cite{Ma08}.

This paper is organized as follows. We first give some general
discussions on anisotropic flows and dihadron correlations in
Sec.~\ref{formula}. We then show in Sec.~\ref{vn} the results on the
centrality dependence of anisotropic flows and their transverse
momentum dependence in most central and mid-central collisions and
in Sec.~\ref{corr} those on the dihadron correlations in most
central collisions. A summary is then given in Sec.~\ref{summary}.

\section{anisotropic flows and dihadron
correlations} \label{formula}

To help understand the results of anisotropic flows and dihadron
correlations in the following sections, we first discuss generally
the contribution from anisotropic flows and their fluctuations as
well as the nonflow contribution to the two-dimensional dihadron
correlation between two particles.

The momentum distribution of produced particles in a heavy ion
collision event can be generally written as
\begin{equation}
\label{f} f(p_T,\phi,\eta) =
\frac{N(p_T,\eta)}{2\pi}\left\{1+2\sum_n
v_n(p_T,\eta)\cos[n(\phi-\Psi_n)]\right\},
\end{equation}
where $\phi$ is the azimuthal angle, $\Psi_n$ is the $n$th-order
event plane angle, and $N(p_T,\eta)$ and $v_n(p_T,\eta)$ are the
number of particles of transverse momentum $p_T$ and pseudorapidity
$\eta$ and their $n$th-order anisotropic flows, respectively. The
correlation between two particles with transverse momenta $p_T^a$
and $p_T^b$ and azimuthal angular difference $\Delta\phi$ and
pseudorapidity difference $\Delta\eta$ within the same event can be
calculated from Eq.~(\ref{f}) by using the orthonormal relation of
the harmonic terms, i.e.,
%\begin{widetext}
\begin{eqnarray}\label{twoparcorr}
&&\frac{d^2N_{\rm pair}^{\rm same}}{d\Delta\eta d\Delta\phi} =
\frac{1}{\eta_{\rm max}-\eta_{\rm min}} \int_{\eta_{\rm min}}^{\eta_{\rm max}}d\eta\notag\\
&\times&\frac{1}{2\pi}\int_0^{2\pi} d\phi  f(p_T^a,\phi,\eta) f(p_T^b,\phi+\Delta\phi,\eta+\Delta\eta)\notag\\
&=&\frac{1}{\eta_{\rm max}-\eta_{\rm min}} \int_{\eta_{\rm min}}^{\eta_{\rm max}}d\eta
\frac{N(p_T^a,\eta)N(p_T^b,\eta+\Delta\eta)}{(2\pi)^2}\notag\\
&\times&\left[1+2\sum_n v_n(p_T^a,\eta)v_n(p_T^b,\eta+\Delta\eta)\cos(n\Delta\phi)\right].
\end{eqnarray}
By neglecting the correlation between particle number and flow, the
event average of the two-particle correlation can then be
approximated by
\begin{eqnarray}\label{tpcsame}
&&\left\langle \frac{d^2N_{\rm pair}^{\rm same}}{d\Delta\eta d\Delta\phi}\right\rangle_{\rm e}\notag\\
&\approx&\frac{1}{\eta_{\rm max}-\eta_{\rm min}}\int_{\eta_{\rm
min}}^{\eta_{\rm max}}d\eta \left\langle\frac{
N(p_T^a,\eta)N(p_T^b,\eta+\Delta\eta)}
{(2\pi)^2} \right\rangle_e \notag\\
&\times&\left[1+2\sum_n \langle v_n(p_T^a,\eta)
v_n(p_T^b,\eta+\Delta\eta) \rangle_{\rm e} \cos(n\Delta\phi)\right],
\end{eqnarray}
where $\langle \cdots \rangle_{\rm e}$ denotes average over all
events.

Assuming that the dependence of the anisotropic flows on the
pseudorapidity is weak, the second term in Eq.~(\ref{tpcsame}) can
be taken out of the integration and further expressed approximately
as
\begin{eqnarray}\label{flow}
&&\langle v_n(p_T^a,\eta) v_n(p_T^b,\eta+\Delta\eta) \rangle_{\rm e}
\cos(n\Delta\phi)\notag\\
&\approx& \langle v_n(p_T^a) \rangle_{\rm e} \langle
v_n(p_T^b) \rangle_{\rm e}\cos(n\Delta\phi)\notag\\
&+& FF(v_n(p_T^a),v_n(p_T^b))\cos(n\Delta\phi) +
NF(\Delta\phi,\Delta\eta),
\end{eqnarray}
where $FF(v_n(p_T^a),v_n(p_T^b))$ is the flow fluctuation
contribution and $NF(\Delta\phi,\Delta\eta)$ is the non-flow
contribution due to jet correlations or resonance decays that are
important only for small $\Delta\eta$ and $\Delta\phi \sim 0$ or
$\pi$. For the case of $p_T^a=p_T^b$, the usual anisotropic flows
calculated using the two-particle cumulant method~\cite{Wan91,Bor01}
after integrating the pseudorapidity difference is then obtained,
i.e.,
\begin{equation}
v_n\{2\} = \sqrt{\langle \cos(n\Delta\phi) \rangle}.
\label{two}
\end{equation}

Using the fact that the first term in Eq.~(\ref{tpcsame}) can be
calculated from the mix-event correlation, i.e.,
\begin{eqnarray}\label{tpcmix}
&&\left\langle \frac{d^2N_{\rm pair}^{\rm mix}}{d\Delta\eta d\Delta\phi}\right\rangle_{\rm e}\notag\\
&=& \frac{1}{\eta_{\rm max}-\eta_{\rm min}}\int_{\eta_{\rm
min}}^{\eta_{\rm max}}d\eta \left\langle
\frac{N(p_T^a,\eta)N(p_T^b,\eta+\Delta\eta)}{(2\pi)^2}\right\rangle_{\rm e},\notag\\
\end{eqnarray}
since the harmonic terms vanish due to the independent event plane
angles in different events, it follows that
\begin{eqnarray}\label{decom}
&&\left\langle \frac{d^2N_{\rm pair}^{\rm same}}{d\Delta\eta
d\Delta\phi}\right\rangle_{\rm e}
/\left\langle \frac{d^2N_{\rm pair}^{\rm mix}}{d\Delta\eta d\Delta\phi}\right\rangle_{\rm e}\notag\\
&&\approx 1+ \langle v_n(p_T^a) \rangle_e \langle
v_n(p_T^b)\rangle_e\cos(n\Delta\phi) \notag\\
&&+FF(v_n(p_T^a),v_n(p_T^b))\cos(n\Delta\phi)+NF(\Delta\phi,\Delta\eta).\notag\\
\end{eqnarray}
The dihadron correlations thus have contributions from anisotropic flows and their fluctuations as well
as the nonflow contribution.

\section{Results}
\subsection{Anisotropic flows}\label{vn}

\begin{figure}[h]
\centerline{\includegraphics[scale=0.8]{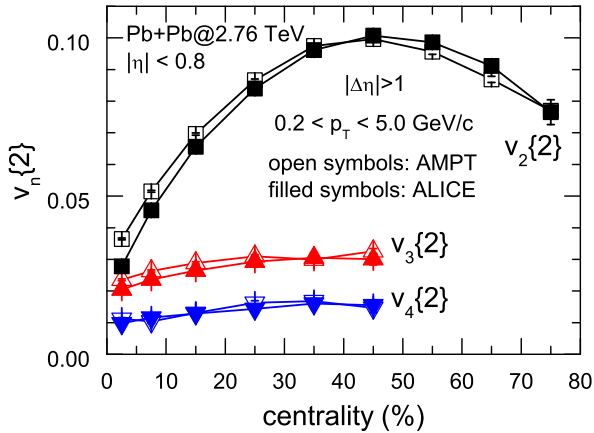}} \caption{(Color
online) Centrality dependence of $v_n (n=2,3,4)$ for
mid-pseudorapidity ($|\eta|<0.8$) charged particles obtained from
the two-particle cumulant method in Pb-Pb collisions at
$\sqrt{s_{NN}^{}}=2.76$ TeV from the string melting AMPT model. The
ALICE data (filled symbols) are taken from Ref.~\cite{Aam11}.}
\label{vncen}
\end{figure}

We first consider the centrality dependence of anisotropic flows
$v_n (n=2,3,4)$ for mid-pseudorapidity charged particles in Pb-Pb
collisions at $\sqrt{s_{NN}^{}}=2.76$ TeV from the string melting
AMPT model using the two-particle cumulant method (Eq.~(\ref{two})).
To reduce nonflow effects on the calculated $v_n$, only pairs with
pseudorapidity difference $|\Delta\eta|>1$ are considered in the
present study as in experimental analyses~\cite{Aam11} . Results for
transverse-momentum-integrated ($0.2<p_T<5.0$ GeV/c) anisotropic
flows are shown by open symbols in Fig.~\ref{vncen}. Compared with
the ALICE data (filled symbols)~\cite{Aam11}, the AMPT model
reproduces reasonably well the centrality dependence of
$v_n(n=2,3,4)$ except that they are slightly larger at small
centralities. This is due to the fact that we have chosen in our
previous work~\cite{Xu11b} to fit the elliptic flow at the $40-50\%$
centrality window.

\begin{figure}[h]
\centerline{\includegraphics[scale=0.8]{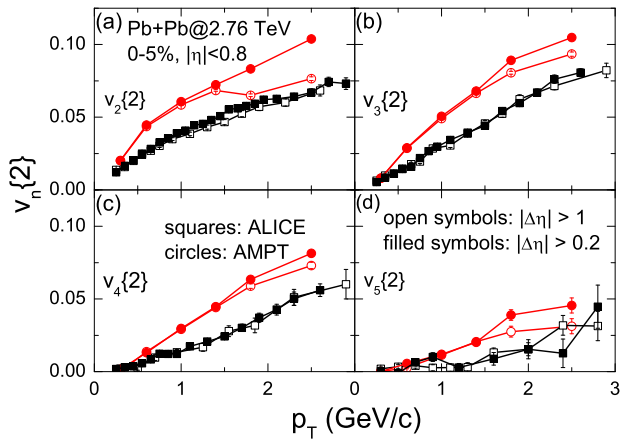}} \caption{(Color
online) Transverse momentum dependence of $v_2$ (a), $v_3$ (b),
$v_4$ (c), and $v_5$ (d) for mid-pseudorapidity ($|\eta|<0.8$)
charged particles obtained from the two-particle cumulant method in
most central $(0-5\%)$ Pb-Pb collisions at $\sqrt{s_{NN}^{}}=2.76$
TeV from the string melting AMPT model. The ALICE data (squares) are
taken from Ref.~\cite{Aam11}. Results for $|\Delta\eta|>0.2$ and
$|\Delta\eta|>1$ are shown by filled and open symbols,
respectively.} \label{vnpt05}
\end{figure}

\begin{figure}[h]
\centerline{\includegraphics[scale=0.8]{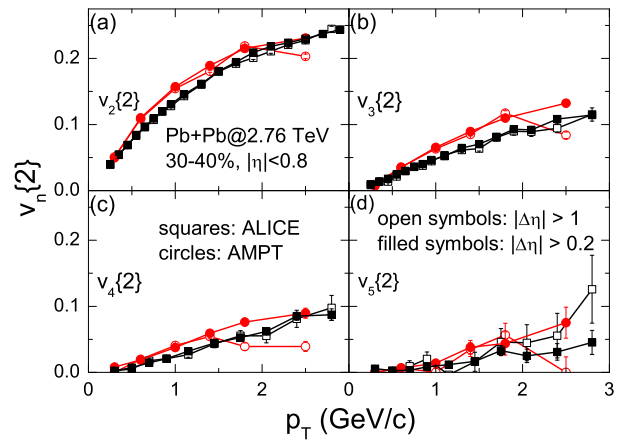}} \caption{(Color
online) Same as Fig.~\ref{vnpt05} but for centrality $30-40\%$.}
\label{vnpt3040}
\end{figure}

In Figs.~\ref{vnpt05} and \ref{vnpt3040}, we display by circles the
transverse momentum  dependence of anisotropic flows
$v_n(n=2,3,4,5)$ of mid-pseudorapidity ($|\eta|<0.8$) charged
particles in the most central $(0-5\%)$ and the mid-central
$(30-40\%)$ Pb-Pb collisions at $\sqrt{s_{NN}^{}}=2.76$ TeV,
respectively, from the AMPT model with string melting, and compare
them with the ALICE data (squares) from Ref.~\cite{Aam11}. It is
seen that the anisotropic flows for different pseudorapidity cut
$|\Delta\eta|>0.2$ and $|\Delta\eta|>1$ are similar to those from
the ALICE data, while at larger transverse momenta in the case of
the smaller pseudorapdity cut the AMPT model gives larger values due
to overestimated nonflow effects. As for RHIC~\cite{Xu11a},
anisotropic flows for the most central collisions from the AMPT
model are overestimated in comparison with those from the ALICE
data, while they are consistent with the ALICE data for mid-central
collisions. The reason for this might be due to the constant parton
scattering cross section used in the AMPT model for all
centralities. Since the temperature of produced partonic matter is
higher in central collisions than in mid-central collisions, a
smaller strong coupling constant and a larger screening
mass~\cite{Kac05a,Kac05b} should be used in calculating the parton
scattering cross section. This is expected to lead to a smaller
parton scattering cross section and smaller anisotropic flows for
central collisions. We note that once the elliptic flow is fitted,
the higher-order flows are automatically consistent with the
experimental data. This indicates that the HIJING model, as an
initial distribution generator, provides the correct initial spatial
anisotropies through the Glauber model calculation.

\subsection{Dihadron correlations}\label{corr}

As in Ref.~\cite{Cha11}, we calculate the dihadron correlation according to
\begin{equation}\label{dih}
\frac{1}{N^{\rm trig}}\frac{d^2N_{\rm pair}}{d\Delta\eta d\Delta \phi} = B(0,0) \times
\frac{S(\Delta\eta,\Delta\phi)}{B(\Delta\eta,\Delta\phi)},
\end{equation}
where
\begin{equation}
S(\Delta\eta,\Delta\phi) =
\frac{1}{N^{\rm trig}}\frac{d^2N_{\rm pair}^{\rm same}}{d\Delta\eta d\Delta\phi}
\end{equation}
is the raw correlation per trigger particle from pairs in same events,
\begin{equation}
B(\Delta\eta,\Delta\phi) =
\frac{1}{N^{\rm trig}}\frac{d^2N_{\rm pair}^{\rm mix}}{d\Delta\eta d\Delta \phi}
\end{equation}
is the background correlation per trigger particle from pairs in
different events, and $B(\Delta\eta=0,\Delta\phi=0)$ is the
normalization factor. Eq.~(\ref{dih}) is thus similar to
Eq.~(\ref{decom}) in Sec.~\ref{formula}.

\begin{figure}[h]
\centerline{\includegraphics[scale=0.9]{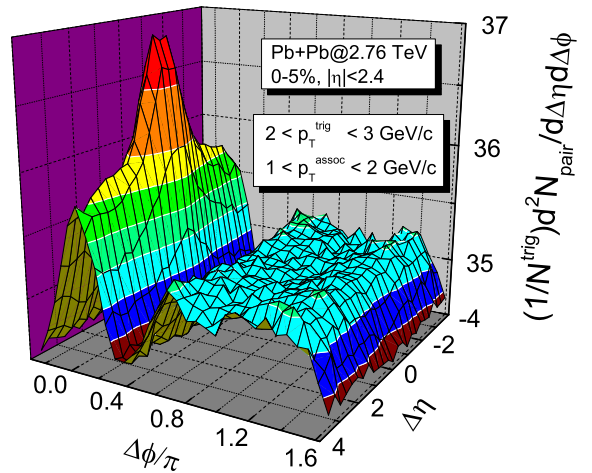}} \caption{(Color
online) Two-dimensional dihadron correlation per trigger particle as
a function of $\Delta \eta$ and $\Delta \phi$ for $1<p_T^{\rm assoc}<2$
GeV/c and $2<p_T^{\rm trig}<3$ GeV/c for $0-5\%$ most central Pb-Pb
collisions at $\sqrt{s_{NN}^{}}=2.76$ TeV from the string melting
AMPT model.} \label{2Dcorr}
\end{figure}

\begin{figure}[h]
\centerline{\includegraphics[scale=0.9]{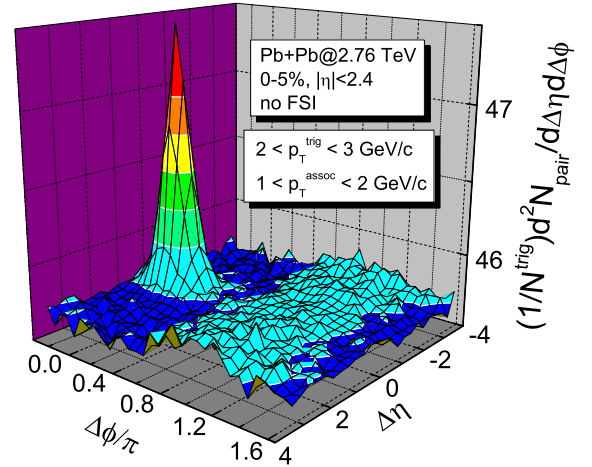}} \caption{(Color
online) Same as Fig.~\ref{2Dcorr} but without final-state
interactions (FSI).} \label{2DcorrnoFSI}
\end{figure}

In Fig.~\ref{2Dcorr} we show the two-dimensional dihadron
correlation per trigger particle from the string melting AMPT model
in the $0-5\%$ most central Pb-Pb collisions at
$\sqrt{s_{NN}^{}}=2.76$ TeV for trigger particles and associated
particles in the transverse momentum windows $2<p_T^{\rm trig}<3$
GeV/c and $1<p_T^{\rm assoc}<2$ GeV/c, respectively. Similar to the
experimental results in Ref.~\cite{Cha11}, there is a peak at
$\Delta\eta=0$ and a ridge structure extending to $|\Delta\eta|=4$
at the near side ($\Delta\phi \sim 0$) as well as a broad structure
at the away side which also extends to $|\Delta\eta|=4$. The ridge
structure was first discovered in Ref.~\cite{Ada06}. Many
explanations have since been proposed
~\cite{Arm04,Sat05,Chi05,Won07,Rom07,Maj07,Shu07} until it was
recently realized that the ridge was dominated by the anisotropic
flows~\cite{Luz11,Aam11,Cha11} as can be seen from
Eq.~(\ref{decom}). To illustrate this effect, we repeat the
calculation by turning off both partonic and hadronic scatterings in
the AMPT model, and the results are shown in Fig.~\ref{2DcorrnoFSI}.
It is seen that the near-side ridge now disappears and the away-side
broad structure is also largely weakened as both are results of the
collective flow that is generated by final-state interactions (FSI).
Furthermore, the peak along the $\Delta\eta$ direction is sharper
than that in the case with FSI. We note that the wider nonflow
contribution in $\Delta\eta$ direction in the case with FSI is
related to the early explanations of the ridge
structure~\cite{Arm04,Sat05,Chi05,Won07,Rom07,Maj07,Shu07}.

\begin{figure}[h]
\centerline{\includegraphics[scale=0.8]{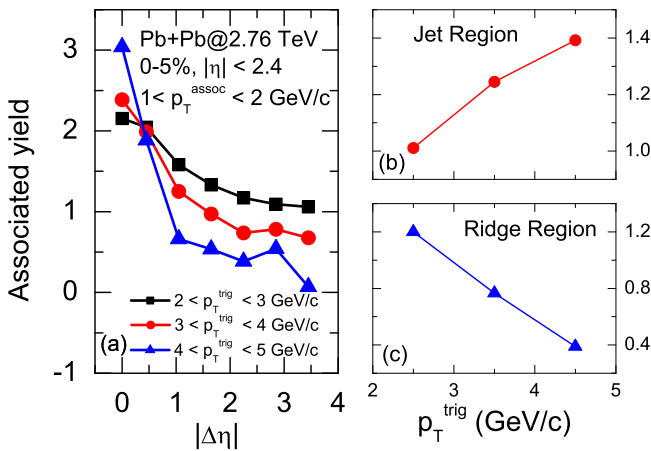}} \caption{(Color
online) The near-side associated yield as a function of
$|\Delta\eta|$ (a) and those as functions of trigger particle
transverse momentum in the jet region (b) and the ridge region (c)
for $0-5\%$ most central Pb-Pb collisions from the string melting
AMPT model.} \label{yield}
\end{figure}

Above results can be quantified by displaying the $|\Delta\eta|$
dependence of the near-side associated yield. Taking the near side
as $|\Delta\phi|<\Delta\phi_\text{ZYAM}$, with
$\Delta\phi_\text{ZYAM}$ obtained using the zero-yield-at-minimum
(ZYAM) method~\cite{Adl06} to fit the dihadron azimuthal correlation
by a second-order polynomial in the range $0.5<\Delta\phi<1.5$ as in
Ref.~\cite{Cha11}, the results are shown in Fig.~\ref{yield} (a) for
different transverse momentum windows of trigger particles
($2<p_T^{\rm trig}<3$ GeV/c, $3<p_T^{\rm trig}<4$ GeV/c, and
$4<p_T^{\rm trig}<5$ GeV/c) but the same transverse momentum window
of associated particles ($1<p_T^{\rm assoc}<2$ Gev/c). It shows a
similar decreasing trend with increasing $|\Delta\eta|$ as in
Ref.~\cite{Cha11}, and it is broader for the case with lower-$p_T$
trigger particles. The width of the near-side associated yield is
related to the broadening of the jet cone affected by the
longitudinal flow, and it seems that the jet cone from higher-$p_T$
jets is more focus and less affected by the longitudinal flow.

\begin{figure}[h]
\centerline{\includegraphics[scale=0.8]{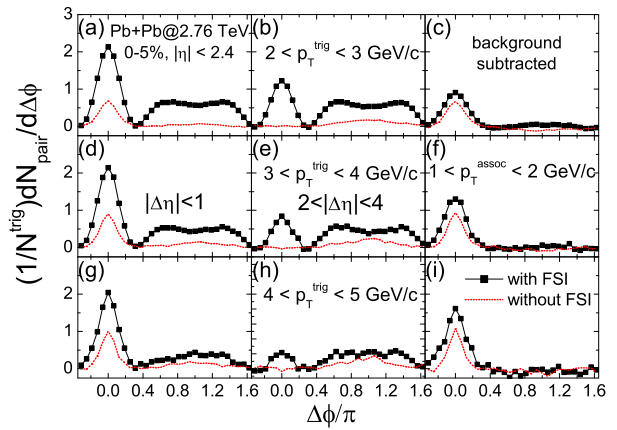}} \caption{(color
online) Short-range ($|\Delta\eta|<1$, (a), (d) and (g)) and
long-range ($2<|\Delta\eta|<4$, (b), (e) and (h)) dihadron
correlations per trigger particle as functions of the azimuthal
angular difference $\Delta\phi$ for different transverse momentum
windows of trigger particles from $0-5\%$ most central Pb-Pb
collisions in the string melting AMPT model with (solid lines) and
without (dashed lines) final-state interactions. The difference
between the short-range and long-range dihadron azimuthal
correlation is shown in panels (c), (f), and (i). } \label{1Dcorr}
\end{figure}

For the dependence of the near-side associated yield on the trigger
particle transverse momentum, we first introduce the short-range
$(|\Delta\eta|<1)$ and long-range $(2<|\Delta\eta|<4)$ dihadron
azimuthal angular correlations by taking average of the
two-dimensional dihadron correlation over the corresponding
$\Delta\eta$ windows~\cite{Cha11} as shown in the left and middle
columns of Fig.~\ref{1Dcorr} for different transverse momenta of
trigger particles. Compared with the short-range correlations, the
long-range correlations have much weaker peaks at the near side due
to the absence of the nonflow contribution. The strength of the
near-side peaks, i.e., the near-side associated yield, decreases
with increasing $p_T$ of trigger particles for both short- and
long-range correlations, although their heights are similar for
short-range correlations. In addition, the strength of away-side
correlations decreases with increasing $p_T$ of trigger particles
for both short-range and long-range correlations.

Results on the trigger particle transverse momentum dependence of
the near-side associated yield  in the jet region and the ridge
region are shown in Fig.~\ref{yield} (b) and (c), respectively. In
the ridge region, i.e., the long-range correlation
($2<|\Delta\eta|<4$), the near-side associated yield decreases with
increasing $p_T$ of trigger particles. This is understandable as the
near-side ridge is dominated by the collective flow that does not
affect much higher-$p_T$ particles. The near-side associated yield
in the jet region, which is the difference between the associated
yield in the long-range correlation ($2<|\Delta\eta|<4$) and the
short-range correlation ($|\Delta\eta|<1$), increases with
increasing $p_T$ of trigger particles. This can be again understood
as higher-$p_T$ particles are more likely to be accompanied by a
larger number of lower-$p_T$ particles because they can fragment
into or dump energy to more lower-$p_T$ particles. All these are
qualitatively consistent with the experimental results~\cite{Cha11}.

Since the dihadron correlation includes contributions from both the
anisotropic flows and the medium response to the back-to-back jet
pairs produced in initial hard collisions, to isolate the latter
effect requires the subtraction of the flow contribution. As
discussed in Refs.~\cite{Luz11,Aam11,Cha11}, the long-range dihadron
correlation is largely caused by the anisotropic flows and their
fluctuations. We can therefore use the long-range dihadron azimuthal
correlation shown in the middle column of Fig.~\ref{1Dcorr} as an
estimation of the background contribution~\cite{Luz11} by neglecting
the pseudorapidity dependence of the anisotropic flows and possible
non-flow contributions for larger $|\Delta\eta|$. The resulting
background-subtracted correlations, i.e., the difference between the
short-range correlations and the long-range correlations, are shown
in Fig.~\ref{1Dcorr} (c), (f), and (i). It is seen that the
resulting away-side correlations are very weak for all cases and
this is similar to the results in Ref.~\cite{Xu11a} obtained using a
different method to estimate the background correlations for central
collisions.  In addition, the near-side correlations in this case
are stronger for higher-$p_T$ trigger particles, which can be
understood as the stronger medium response to higher-$p_T$ jets. For
comparisons, we also show in Fig.~\ref{1Dcorr} by dashed lines the
dihadron azimuthal correlations obtained without final-state
interactions, which are similar to those from p+p collisions at the
same energy. The background-subtracted away-side correlations in
this case is similar to that obtained with final-state interactions.
The smearing of the away-side jet in heavy ion collisions without
final-state interactions or p+p collisions is due to the fairly low
$p_T$ of trigger and associated particles considered here, which can
lead to an away-side jet with very different pseudorapidity from
that of the triggered jet~\cite{xnwang} as shown in the middle
column of Fig.~\ref{1Dcorr}. A peak in the away-side correlations of
such collisions would, however, appear if the associated particles
have similar momenta as those of the trigger particles or the
trigger particles have very high $p_T$.  This is very different from
the case where final-state interactions are included. Because of jet
quenching as a result of final-state interactions, no peak structure
is seen in the away-side correlations unless the trigger particles
have extremely high $p_T$.

\section{Conclusions and discussions}
\label{summary}

We have studied higher-order anisotropic flows and dihadron
correlations in Pb-Pb collisions at $\sqrt{s_{NN}^{}}=2.76$ TeV
within a multiphase transport model with parameters fitted to
reproduce the measured multiplicity density of mid-pseudorapidity
charged particles in central collisions and their elliptic flow in
mid-central collisions in the previous work. We have found that the
resulting higher-order anisotropic flows slightly overestimate the
experimental data at small centralities but are consistent with them
at other centralities. We have obtained the ridge structure along
the pseudorapidity direction in the near side of the dihadron
correlations, and it disappears when final-state interactions are
turned off in our model. We have also studied both the short-range
and long-range dihadron azimuthal correlations for different
transverse momenta of trigger particles, and they are seen to be
quantitatively consistent with experimental results. We have further
attempted to determine the background-subtracted short-range
dihadron azimuthal correlations by taking the long-range dihadron
azimuthal correlations as the background, and they are found to be
similar to those obtained previously using a different method.

\begin{acknowledgments}
We thank the ALICE Collaboration for providing the data on
anisotropic flows. This work was supported in part by the U.S.
National Science Foundation under Grants No. PHY-0758115 and No.
PHY-1068572, the US Department of Energy under Contract No.
DE-FG02-10ER41682, and the Welch Foundation under Grant No. A-1358.
\end{acknowledgments}

\end{document}